\PassOptionsToPackage{table,xcdraw}{xcolor}
\documentclass[sigconf,nonacm]{acmart}

\newcommand*\rev[1]{#1} 

\newcommand{\changedimage}[1]{#1}

\usepackage{hyperref}
\usepackage{cleveref}
\usepackage{caption}
\usepackage{graphicx}
\usepackage{subcaption}
\usepackage{pifont}
\usepackage{multicol}
\usepackage[normalem]{ulem}

\usepackage{tikz}
\newcommand*\circled[1]{\tikz[baseline=(char.base)]{
            \node[shape=circle,draw,inner sep=0.5pt] (char) {#1};}}
\newcommand{\systemname}{Redbench}
\newcommand{\cmark}{\ding{51}}

\usepackage{stackengine}

\usepackage{calc}  

\DeclareRobustCommand{\tcircled}[1]{%
  \makebox[\widthof{\scalebox{1.3}{$\bigcirc$}}][c]{%
    \stackinset{c}{}{c}{-0.15ex}{\scalebox{1.3}{$\bigcirc$}}{#1}%
  }%
}

\newcommand{\select}{\texttt{\small SELECT}}
\newcommand{\SELECT}{\select}
\newcommand{\INSERT}{\texttt{\small INSERT}}
\newcommand{\update}{\texttt{\small UPDATE}}
\newcommand{\UPDATE}{\update}
\newcommand{\delete}{\texttt{\small DELETE}}
\newcommand{\DELETE}{\delete}
\newcommand{\CTAS}{\texttt{\small CTAS}}
\newcommand{\MERGE}{\texttt{\small MERGE}}
\newcommand{\ANALYZE}{\texttt{\small ANALYZE}}
\newcommand{\COPY}{\texttt{\small COPY}}

\begin{document}

\title{\systemname{}: Workload Synthesis From Cloud Traces}

\author{Johannes Wehrstein}
\orcid{0000-0002-7152-8959}
\affiliation{%
  \institution{
    Technical University of Darmstadt
  }
}

\author{Roman Heinrich}
\orcid{0000-0001-7321-9562}
\affiliation{%
  \institution{
    Technical University of Darmstadt \& DFKI
  }
}

\author{Mihail Stoian}
\affiliation{%
  \institution{
    University of Technology Nuremberg
  }
}

\author{Skander Krid}
\authornote{Work done while at University of Technology Nuremberg.}
\affiliation{%
  \institution{
    Snowflake
  }
}

\author{Martin Stemmer}
\affiliation{%
  \institution{
    Technical University of Darmstadt
  }
}

\author{Andreas Kipf}
\affiliation{%
  \institution{
    University of Technology Nuremberg
  }
}

\author{Carsten Binnig}
\orcid{0000-0002-2744-7836}
\affiliation{%
  \institution{
    Technical University of Darmstadt \& DFKI \& hessen.AI
  }
}

\author{Muhammad El-Hindi}
\affiliation{%
  \institution{
    Technical University of Munich
  }
}

\begin{abstract}
Workload traces from cloud data warehouse providers reveal that standard benchmarks such as TPC-H and TPC-DS fail to capture key characteristics of real-world workloads, including query repetition and string-heavy queries.
In this paper, we introduce \systemname{}, a novel benchmark featuring a workload generator that reproduces real-world workload characteristics derived from traces released by cloud providers.
\systemname{} integrates multiple workload generation techniques to tailor workloads to specific objectives, transforming existing benchmarks into realistic query streams that preserve intrinsic workload characteristics. 
By focusing on inherent workload signals rather than execution-specific metrics, \systemname{} bridges the gap between synthetic and real workloads. 
Our evaluation shows that (1) \systemname{} produces more realistic and reproducible workloads for cloud data warehouse benchmarking, and (2) \systemname{} reveals the impact of system optimizations across four commercial data warehouse platforms.
We believe that \systemname{} provides a crucial foundation for advancing research on optimization techniques for modern cloud data warehouses.
\end{abstract}

\maketitle


\section{Introduction}

\begin{figure}
    \centering
    \changedimage{\includegraphics[width=.95\linewidth,trim={0cm 0.4cm 0 0.2cm},clip]{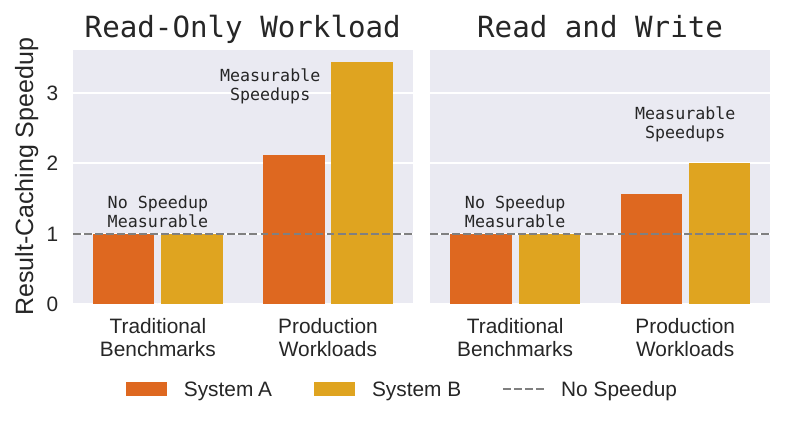}}
    \vspace{-1ex}
    \caption{Result caching speedups for traditional benchmarks versus production workloads. 
    Traditional benchmarks (e.g., JOB, CEB) fail to capture workload-driven effects, such as query repetitions and temporal correlations, that are present in production traces (e.g., Redset) and are essential for evaluating optimizations like result caching.}
    \vspace{-4ex}
    \label{fig:intro_plot}
\end{figure}

\noindent\textbf{The need for new analytical benchmarks.}
Analytical benchmarks such as TPC-H and TPC-DS have been the cornerstones of analyzing the performance of analytical database engines.
However, recent studies~\cite{DBLP:journals/pvldb/SzlangBCDFHOOM25, DBLP:journals/pvldb/RenenHPVDNLSKK24, DBLP:journals/pvldb/RenenL23, DBLP:journals/pvldb/DingCGN21, DBLP:conf/sigmod/VogelsgesangHFK18} show that these classical benchmarks fail to represent characteristics of real-world workloads.
For instance, an analysis of the Amazon Redshift fleet~\cite{DBLP:journals/pvldb/RenenHPVDNLSKK24} found that production workloads are more write-heavy and highly repetitive.
Moreover, they exhibit long-tailed query and table size distributions.
A recent study by Snowflake reached a similar conclusion ~\cite{DBLP:journals/pvldb/SzlangBCDFHOOM25}.

\noindent\textbf{Workload synthesis from real-world traces.}
To address this disparity, we argue that a new benchmark is needed that better reflects the characteristics of production traces released by cloud vendors along with their studies.
These traces capture key workload properties, including query patterns, operator counts, and resource usage, and reveal trends such as mixed read~/~write workloads, bursty activity, and the prevalence of specific operators.
However, because the traces omit both the actual SQL text and the underlying data, these traces cannot serve as end-to-end benchmarks.
This limitation has motivated research on synthesizing replayable workloads that reflect the observed workload characteristics.

\noindent\textbf{\systemname{} at a glance.}
To address this, we propose \systemname{}, extending our previous work~\cite{DBLP:journals/corr/abs-2506-12488}.
\systemname{} takes the workload trace from Amazon Redshift, as well as a schema from an existing benchmark (e.g., JOB or TPC-H), and generates \emph{trace-shaped workloads}. \systemname{} does so by reproducing production properties extracted from a trace: (i) scheduling exact and structural \emph{repetitions} based on query hashes and scanned tables, (ii) preserving \emph{temporal dynamics} (bursts, diurnal cycles, idle gaps), and (iii) interleaving \emph{mixed query types} (\SELECT{} and diverse DML) to model realistic cache invalidations. Concretely, \systemname{} either matches trace entries to existing benchmark queries and reorders them accordingly, or it generates new SQL guided by trace statistics to hit target bytes-read, join shapes, and access patterns. The result is a drop-in workload that retains the robustness of well-known benchmarks while mimicking the properties of production workloads.

\noindent\textbf{Testing under realistic workload characteristics.}
By turning familiar benchmarks into cloud-like, trace-shaped workloads, \systemname{} enables academia and industry to test optimizations under \emph{realistic workload characteristics} without access to proprietary SQL or data.
For example, as shown in \Cref{fig:intro_plot}, standard benchmarks overlook query repetition patterns observed in studies from cloud vendors. As a result, they fail to capture the effects of result caching, a crucial feature of modern cloud data warehouses.
Despite result caching being enabled in two modern analytical engines (System~A and B), query performance does not improve on traditional benchmarks.
In contrast, when evaluated on workloads generated by \systemname{}, query performance can improve by more than 3$\times$.

\noindent\textbf{\systemname{} enables targeted optimizations.}
Overall, \systemname{} is not only a valuable resource for analyzing and comparing system performance, but also a foundation for developing targeted optimizations in modern cloud data warehouses.
We believe \systemname{} significantly lowers the barrier to evaluating new features in cloud data warehouses and can accelerate progress in the field: researchers can now validate their ideas against production-like repetition, temporal, and mixed-type query patterns, enabling faster iteration and more meaningful results.
In particular, \systemname{} provides a practical framework for studying techniques that exploit reuse, such as result and subresult caching~\cite{DBLP:journals/pvldb/GarrodMAMMOT08}, view and materialization reuse~\cite{DBLP:journals/debu/GuptaM95}, and adaptive clustering~\cite{DBLP:journals/debu/GuptaM95,DBLP:conf/sigmod/HilprechtBR20}, workload-aware indexing~\cite{DBLP:conf/icde/BonczCFHL0NSSZ23,DBLP:conf/sigmod/El-HindiZBK16}, and fitted or learned DBMS components~\cite{DBLP:conf/cidr/HilprechtBBEHKR20}, and can inspire new workload-driven optimizations grounded in production-like patterns.

\noindent\textbf{Combining multiple synthesis approaches.}
In this work, we build on our previous benchmark, Redbench-v1~\cite{DBLP:journals/corr/abs-2506-12488}.
Unlike \systemname{}, Redbench-v1 generated synthetic workloads using a simpler approach that naively sampled queries from traditional benchmarks to primarily preserve repetition rates.
While \systemname{} builds on the same core idea, it introduces a new methodology that combines matching-based and generation-based workload synthesis.
This approach enables \systemname{} to more closely reproduce the characteristics of traces released by cloud providers and to expand query coverage by incorporating \INSERT{}, \UPDATE{}, and \DELETE{} statements, which are interleaved with reads and represent a significant portion of real analytical workloads.

\noindent\textbf{Contributions.}
We present \emph{\systemname{}}, a new benchmark and workload generator designed to capture the realistic cloud analytics properties summarized in \Cref{tab:benchmark_overview}.
Our main contributions are:
\begin{itemize}
    \vspace{-1ex}
    \setlength{\parskip}{0pt} 
    \setlength\itemindent{0pt}
    \setlength\leftskip{-15pt}
    \item \textbf{Methodology of synthesis approaches.} In Section~\ref{sec:analysis_of_synthesis}, we introduce a new synthesis methodology used in \systemname{}, which enables it to capture key characteristics of analytical workload traces released by cloud providers.
    \item \textbf{Realistic workload synthesis.} Building on this methodology, we describe the design of \systemname{}.
    \systemname{} takes a support database (for example, an existing database such as IMDb) and generates SQL workloads that incorporate key properties of production workloads.
    It supports both structural and exact query repetitions, as well as temporal patterns such as burstiness, cyclical drifts, and query ordering, along with diverse query types, including \INSERT{}, \UPDATE{}, and \DELETE{} statements interleaved with \SELECT{} queries. \systemname{} also supports string-heavy queries.
    \item \textbf{Evaluation of workload-driven optimizations.} We present a first comprehensive evaluation of \systemname{} workloads on four modern cloud analytics systems and demonstrate that they effectively implement workload-driven database optimizations.
    This establishes a practical foundation for studying and improving cloud data warehouses using workloads that capture \emph{true} production characteristics, thereby accelerating research and practice and fostering the development of new workload-driven optimizations.
    \vspace{-.5ex}
\end{itemize}
Overall, \systemname{} aims to be the \emph{go-to benchmark} for realistic, workload-driven evaluation of cloud database systems.
By explicitly modeling the patterns and characteristics observed in modern analytical workloads, it produces a benchmark that is more faithful and representative than existing alternatives.

\section{Background and Related Work}

In the following sections, we provide the necessary background on the information contained in the recently released workload traces and then discuss other generation-based approaches similar to \systemname{}, explaining why they are insufficient to capture all the characteristics of these traces.
There are similar efforts in related areas~\cite{DBLP:conf/sigmod/AkenDPCC15, DBLP:conf/sigmod/BattleEACSZBFM20, DBLP:journals/pvldb/ZhangJ0MCJNWZL23,DBLP:journals/dbsk/ElHindiAKB22}, however we focus on  OLAP in the following.

\subsection{Workload Traces}
Cloud providers have recently released anonymized workload traces \cite{DBLP:conf/nsdi/VuppalapatiMATM20, DBLP:journals/pvldb/RenenHPVDNLSKK24}, offering the first opportunity to study real analytical workloads at scale.
The most widely used datasets are \emph{Snowset} and \emph{Redset}, each revealing properties absent from synthetic benchmarks.

\noindent\textbf{Snowset.}
Snowset~\cite{DBLP:conf/nsdi/VuppalapatiMATM20} is a publicly released Snowflake workload trace containing statistics for $\approx$69 million queries over two weeks in early 2018.
\rev{Its analyses reveal highly skewed resource usage, with a small fraction of queries consuming most CPU time.
Snowset also contains many read/write (ETL-style) jobs and a balanced operator mix.
For example, only about 50\% of execution time is in scans/filters (vs.\ $\sim$84\% in single-node TPC-H), with the rest in joins, aggregation, and DML.
These findings suggest TPC-H/DS (which primarily assume scans and relatively simple joins) do not reflect real cloud workloads.
Snowflake queries are more varied in type, involve heavier DML, and exhibit dramatic skew across query sizes.}

\noindent\textbf{Redset.}
Redset~\cite{DBLP:journals/pvldb/RenenHPVDNLSKK24} is an Amazon Redshift workload trace consisting of three months of query metadata from 200 instances (both serverless and provisioned).
It likewise reveals extreme skew and long-tail behavior:
Redset queries have substantially longer execution times than typical TPC-DS queries, and a tiny fraction consumes a disproportionate share of resources (e.g., fewer than 0.1\% of queries account for roughly 25\% of total CPU time).
In contrast, TPC-DS workloads are comparatively uniform.

\subsection{Workload Synthesis}
\label{subsec:related_wl_synth}
Motivated by the mismatch between synthetic benchmarks and production traces, several works have proposed generating workloads that better reflect real-world workloads.
Similar to \systemname{}, these approaches typically draw from pools of benchmark queries (e.g., TPC-H/DS, JOB) and use trace statistics to approximate the cost and structure of real workloads.

\noindent\textbf{Stitcher.}
Stitcher~\cite{DBLP:conf/edbt/WanZC0LLTSCKK23} is a learned synthesizer that combines baseline benchmarks (e.g., TPC-C, TPC-DS) to match a target performance profile.
It uses Bayesian optimization to tune query frequencies and concurrency.
However, Stitcher primarily fits aggregate metrics (CPU, I/O, latency) and does not ensure that the synthetic workload matches the target operator mix (e.g., joins vs.\ scans).
Moreover, Stitcher selects from the full query pool without fine-grained filtering, which limits its ability to tailor the query mix to specific workload characteristics.

\noindent\textbf{Cloud Analytics Benchmark.}
The Cloud Analytics Benchmark (CAB)~\cite{DBLP:journals/pvldb/RenenL23} builds upon Snowset analyses, retaining the TPC-H schema and queries while incorporating cloud-specific aspects, such as elasticity and multi-tenancy.
It generates multiple database ``tenants'' of varying sizes and assigns query streams to each, with time-varying arrival rates modeled after Snowset patterns.
However, query selection within each stream is largely heuristic: CAB's generator ``randomly selects queries from the database's query pool''.
In effect, CAB samples from fixed query templates rather than calibrating the workload to actual trace statistics.
Thus, while CAB reflects cloud behaviors like burstiness and tenant mix, it fails to reproduce real workload cost or operator distributions.

\noindent\textbf{DriftBench.}
\citeauthor{DBLP:journals/corr/abs-2510-10858} introduced DriftBench~\cite{DBLP:journals/corr/abs-2510-10858}, a benchmark modeling and generating both data and workload drift.
It formalizes separate classes of data and workload drift (e.g., cardinality, predicate or structural changes) with predefined temporal patterns to evaluate drift-aware components such as cardinality estimators.
However, its drifts remain synthetic and regular, lacking the irregular, workload-specific drift of real systems


\noindent\textbf{SQLBarber.}
SQLBarber~\cite{DBLP:journals/corr/abs-2507-06192,DBLP:conf/sigmod/LaoT25} uses large language models (LLMs) to generate SQL queries from high-level specifications, accepting natural-language constraints and target cost distributions.
Leveraging execution statistics from Snowflake and Redshift, SQLBarber can produce large volumes of queries whose cost profiles match real-world characteristics.
As a fully generative system, it creates novel SQL patterns but does not aim to replay historical workloads or explicitly model workload dynamics.
It fits a desired cost histogram rather than reconstructing temporal or multi-query behavior, and thus complements synthesis methods rather than reproducing an observed workload end-to-end.

\noindent\textbf{PBench.}
\citeauthor{DBLP:journals/pvldb/ZhouLUWMZZPHLPKMF25} introduce PBench~\cite{DBLP:journals/pvldb/ZhouLUWMZZPHLPKMF25}, which constructs synthetic workloads from existing benchmarks using real-world statistics as targets.
Given execution summaries (e.g., CPU time, operator counts) from traces such as Snowset or Redset, PBench formulates a multi-objective optimization to select and combine queries to match target cost and operator distribution, and employs an LLM to generate queries when needed.
\rev{While this advances beyond heuristic sampling (as in CAB) or metric-only matching (as in Stitcher), PBench still operates on aggregate statistics and therefore only partially models \emph{temporal dynamics} (diurnal patterns, bursts, drifts),\footnote{\rev{Depending on the specified window length.}} while not having explicit support for neither \emph{query repetitions} nor DMLs.}
Moreover, despite LLM support, control over query features (e.g., string-heavy predicates or semi-structured access) remains limited.
In summary, PBench narrows the gap by matching cost and operator statistics but falls short of reproducing the structural and temporal properties of production cloud workloads.

\noindent\textbf{SQLStorm.}
\citeauthor{DBLP:journals/pvldb/SchmidtLBN25} introduced SQLStorm~\cite{DBLP:journals/pvldb/SchmidtLBN25}, an LLM-based benchmark generation system that generates SQL workloads over a real-world dataset.
\rev{SQLStorm generates a large query set covering many advanced SQL constructs (e.g., complex joins, window functions, JSON, and string predicates), resulting in a long-tailed distribution of query complexity and operator counts (e.g., ``up to 38 operators'' \rev{per query}).}
Consequently, it can challenge database engines beyond the predictable behavior of static, template-based benchmarks.
However, like other generation-centric approaches, SQLStorm does not explicitly model query repetitions or temporal workload patterns, such as diurnal cycles or burstiness.
Thus, while SQLStorm advances SQL diversity and stress-testing capabilities, it does not capture workload dynamics or repeated query behavior necessary for evaluating workload-driven optimizations.

\begin{table}[t]
    \caption{Existing classical and synthesized benchmarks~\cite{tpc_ds, tpc_h,DBLP:journals/pvldb/LeisGMBK015,DBLP:journals/corr/abs-2507-07471,DBLP:journals/pvldb/SchmidtLBN25,DBLP:journals/pvldb/RenenL23,DBLP:journals/pvldb/ZhouLUWMZZPHLPKMF25} ignore critical production-workload features required to effectively evaluate workload-driven optimizations. Some only use generated (Gen) instead of real data.}
    \label{tab:benchmark_overview}
    \centering
    \vspace{-2ex}
    \scalebox{0.91}{
    \begin{tabular}{l|ccccc}
    \hline
\textbf{Benchmark} & \textbf{\begin{tabular}[c]{@{}c@{}}String-\\ Heavy \end{tabular}} & \textbf{\begin{tabular}[c]{@{}c@{}}Temporal \\ Patterns\end{tabular}} & \textbf{\begin{tabular}[c]{@{}c@{}}Repeti-\\ tions\end{tabular}} & \textbf{Writes} & \textbf{Data} \\ \hline
\rowcolor{gray!10} 
TPC-H / DS &  &  &  &  & Gen \\
JOB(-Complex) & \cmark &  &  &  & Real \\
\rowcolor{gray!10} 
SQLStorm & \cmark &  &  &  & Real \\
SQLBarber & \cmark &  &  &  & Real \\
\rowcolor{gray!10} 
Stitcher &  & \cmark & &  & Gen \\ 
DriftBench &  & \cmark  &  &  & Real \\
\rowcolor{gray!10} 
CAB &  & \cmark &  &  & Gen \\
PBench & (\cmark) & \cmark &  &  & Real \\ \hline
\rowcolor{gray!10} 
\textbf{\systemname} & \cmark & \cmark & \cmark & \cmark & Real \\ \hline
\end{tabular}
}
\vspace{-2ex}
\end{table}

\subsection{Limitations of Existing Approaches}
Although proposals such as CAB, Stitcher, PBench, and SQLBarber improve upon traditional benchmarks, they primarily focus on matching aggregate statistics (e.g., operator distributions, runtimes) and fail to capture several challenging workload properties observed in production systems (cf.~\Cref{tab:benchmark_overview}):
\begin{itemize}
    \setlength\itemindent{0pt}
    \setlength\leftskip{-18pt}
    \item \textbf{String-heavy queries.} Real workloads include complex predicates on strings, JSON, and nested data, which are largely absent from standard benchmarks such as TPC-H.
    \item \textbf{Temporal patterns.} Workloads exhibit diurnal and weekly cycles, sudden spikes, and long-term drifts. Snowset~\cite{DBLP:conf/nsdi/VuppalapatiMATM20}, for instance, shows oscillations with more read-only queries during business hours, while Redset~\cite{DBLP:journals/pvldb/RenenHPVDNLSKK24} varies significantly over time.
    Existing generators do not model such time-based patterns, and any temporal variation typically arises only by chance.
    \item \textbf{Repetitions.} Real workloads often reissue identical or structurally similar queries repeatedly.
        Existing generators may incidentally replay queries multiple times, but they lack explicit mechanisms to control the number and type of repetitions.
    \item \textbf{Writes.} Modern analytics workloads mix reads with write-heavy data-loading or transformation queries (\INSERT{}, \UPDATE{}, \DELETE{}, \CTAS{}, \MERGE{}, etc.).
        Most benchmark-driven synthesis methods are \SELECT{}-heavy and omit these operations (e.g., CAB and Stitcher derive from TPC-H, which contains no \INSERT{} or \CTAS{} queries).
\end{itemize}
As shown in \Cref{tab:benchmark_overview}, \systemname{} is the only approach to cover all these aspects.

\begin{figure}[t]
    \centering
    \includegraphics[width=0.99\linewidth,trim={0cm 0.0cm 0 0.1cm},clip]{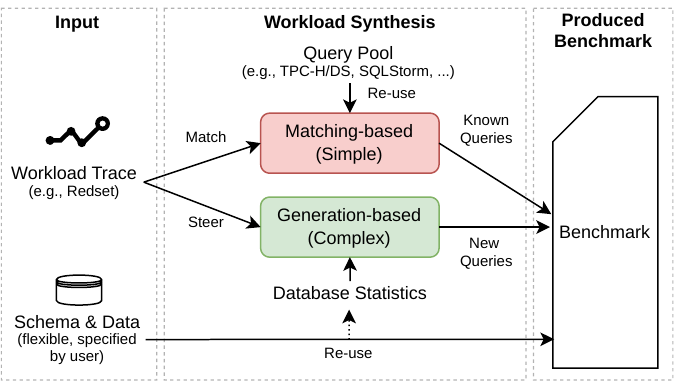}
    \vspace{-2ex}
    \caption{\systemname{} uses a given workload trace and data set as a basis for workload synthesis. Currently, \systemname{} supports two synthesis approaches that users can utilize to generate benchmarks with trace-shaped queries.}
    \label{fig:methodology}
    \vspace{-4ex}
\end{figure}

\section{\systemname{} Methodology \& Overview}
\label{sec:analysis_of_synthesis}


While \systemname{}'s naming has been inspired by the Redset, it follows a Redset-independent methodology to generate \emph{trace-shaped} benchmarks with the challenging properties of analytical production workloads.
As depicted in \Cref{fig:methodology}, the key ingredients in \systemname{} are a workload trace and a dataset with its table schema.
These two inputs form the basis for the workload synthesis.

\noindent\textbf{Trace-shape synthesis.}
\systemname{} assumes that the workload trace records anonymized structural and runtime metrics per query, such as query type, arrival timestamp, join count, accessed tables, and a query hash (cf. \Cref{tab:redset}).
These statistics and features steer \systemname{}'s workload synthesis and enable it to infer query frequencies, complexities, scansets, repetitions, and temporal locality.
\rev{For instance, \systemname{} exploits timestamps to determine the order of queries, which is crucial for capturing effects like cache invalidations from DML queries.
Similarly, when \systemname{} observes the same query hash (i.e., identical SQL fingerprint) and scanset (Read Tbls), it infers a query repetition.}

\noindent\textbf{Repetition types.}
Repetitions can occur in two types: full query repetitions and scanset repetitions.
Full query repetitions describe the exact reoccurrence of the same textual SQL query.
For such repetitions, caching the output result of the query and returning it when the query recurs enables skipping query execution entirely.
In contrast, scanset repetitions describe a structural similarity among the queries, i.e., they operate on the same tables.
However, they can differ in filters, join conditions, or aggregations, making scanset repetitions relevant for intermediate materialization or predicate caching.
Furthermore, we distinguish between DML-aware and non-DML-aware repetitions.
Non-DML-aware repetitions ignore interleaved writes that, in reality, might invalidate cached results (e.g., \texttt{rrwrr} counts as 3 repetitions).
DML-aware repetitions account for interleaved writes to previously queried tables (e.g., \texttt{rrwrr} counts as two different repetitions).

\noindent\textbf{Query-level synthesis.}
Since the original SQL text is not included in a workload trace, \systemname{} relies on workload synthesis to generate SQL queries.
Unlike PBench, which uses aggregate statistics in a given interval to synthesize a set of queries, our approach performs query-level synthesis to match the statistics and properties of a given query (e.g., number of joins, bytes read, etc.).
\systemname{} comes with two synthesis modes as depicted in \Cref{fig:methodology} (middle):
\begin{itemize}
    \setlength\itemindent{0pt}
    \setlength\leftskip{-18pt}
    \item \textbf{Matching-based:} \rev{Reuses queries from an existing query pool (e.g., human-designed benchmarks such as TPC-H/DS, or optionally LLM-generated pools such as SQLStorm) and parametrizes them to match repetitions, scansets, and query runtimes.}
    \item \textbf{Generation-based:} Generates SQL from dataset statistics to match structural and runtime targets (joins, scansets, bytes read).
\end{itemize}

\noindent\textbf{Matching- vs. generation-based synthesis.}
\rev{\systemname{} enables users to select the synthesis approach based on their benchmarking goals, as summarized in \Cref{tab:matching_vs_generation}.
The matching-based approach is simpler and relies on stable, user-selected queries from established benchmarks (e.g., JOB, TPC-H), reducing the likelihood of system failures.
This battle-tested pool provides known SQL feature compatibility, making it well-suited for environments supporting only a restricted SQL subset, common for early-stage research prototypes and operator-limited systems.
By restricting execution to a user-selected query set, matching-based synthesis enables fast iteration and short time-to-feedback cycles when evaluating ideas like caching or query optimization, as observed when this Redbench variant was used to evaluate index advisors~\cite{robust-index}.
However, because it is constrained by the source benchmark's query set, it cannot reproduce all statistical characteristics of real-world workloads (\Cref{subsec:synthesis_quality}).
Once a prototype matures, we therefore recommend switching to \systemname{}'s generation-based variant, which goes beyond a fixed query pool.
In contrast, the generation-based approach provides higher fidelity and flexibility by constructing entirely new SQL queries that are statistically steered to match observed workload properties, including bytes-read distributions, scan sets, and realistic read/write mixes.
By synthesizing queries from scratch, it exercises a broader range of execution paths and stress-tests the DBMS more thoroughly, enabling trace-faithful benchmarking that more closely reflects production behavior.
However, this added flexibility comes at the cost of higher complexity and greater risk of system failures (e.g., out-of-memory errors due to large intermediate results), adding additional strain to the system.}

\begin{table}[t]
\caption{A simplified example of Redset input data. Query 2 is a repetition of Query 0 (same hash and scanset). However, the `\INSERT{}` at time 6pm (query 1) modifies table `1`, potentially invalidating a cached result for query 2.}
\label{tab:redset}
\centering
\vspace{-1ex}
\scalebox{0.9}{
\begin{tabular}{lllllllll}
\hline
\# & Type&Time&Joins & \begin{tabular}[c]{@{}l@{}}Read \\ Tbls\end{tabular} & \begin{tabular}[c]{@{}l@{}}Write\\ Tbl\end{tabular} & \begin{tabular}[c]{@{}l@{}}Run-\\time\end{tabular} & \begin{tabular}[c]{@{}l@{}}Bytes \\ Read\end{tabular} & Hash \\ \hline
0 & select & 4pm & 1 & 0,1 & - & 820 ms & 9 M & a62e \\
1 & insert & 6pm & 1 & 0,3 & 1 & 420 ms & 4 M & 73b9 \\
2 & select & 7pm & 1 & 0,1 & - & 750 ms & 10 M & a62e \\
3 & select & 9pm & 0 & 1 & - & 200 ms & 78 M & 962c \\ \hline
\end{tabular}
}
\vspace{-3ex}
\end{table}
\noindent\textbf{Dataset and schema.}
\systemname{} supports different datasets and schemas for workload synthesis.
The dataset controls common properties such as data distribution and correlation, while the schema constrains synthesis: \rev{too few tables prevent queries joining many tables, and even with sufficient tables, \systemname{} must address key questions such as schema matching (which tables to use for a query's scanset) and accurately reflect scanned bytes---questions discussed in the following section.}

\noindent\textbf{Output and benchmarking.}
As \Cref{fig:methodology} shows, \systemname{} outputs a replayable workload trace consisting of SQL queries and their corresponding arrival timestamps (as determined by the input trace).
To run the benchmark, users load the provided dataset and execute the queries against a desired database engine.
\begin{table}[t]
    \centering
    \caption{Comparison of the matching-based and generation-based synthesis approaches.}
    \label{tab:matching_vs_generation}
    \vspace{-2ex}
    \scalebox{0.87}{
    \begin{tabular}{l|ll}
\hline
\textbf{} & \textbf{Matching-Based} & \textbf{Generation-Based} \\ \hline
\rowcolor{gray!10} 
\textbf{\begin{tabular}[c]{@{}l@{}}Query\\ Source\end{tabular}} & \begin{tabular}[c]{@{}l@{}}Reuses queries from\\ existing benchmarks\end{tabular} & \begin{tabular}[c]{@{}l@{}}Generates novel \\ queries\end{tabular} \\
\textbf{\begin{tabular}[c]{@{}l@{}}Query \\ Types\end{tabular}} & \begin{tabular}[c]{@{}l@{}}\rev{\SELECT{}, only} \\ \rev{single-row re-inserts}\end{tabular} & \begin{tabular}[c]{@{}l@{}}\SELECT{}, \INSERT{}, \\ \UPDATE{}, \DELETE{}\end{tabular} \\
\rowcolor{gray!10} 
\textbf{Complexity} & \begin{tabular}[c]{@{}l@{}}Limited by source \\ benchmark (e.g., TPC-H)\end{tabular} & \begin{tabular}[c]{@{}l@{}}High flexibility, \\ complex structures\end{tabular} \\
\textbf{Repetition} & Exact repetitions & \begin{tabular}[c]{@{}l@{}}Structural \& \\ exact repetitions\end{tabular} \\
\rowcolor{gray!10} 
\textbf{Fidelity} & \begin{tabular}[c]{@{}l@{}}Matches num joins, \\ repetitions\end{tabular} & \begin{tabular}[c]{@{}l@{}}Matches scansets, \\ joins, bytes read, etc.\end{tabular} \\ 
\textbf{\rev{When to use}} & \begin{tabular}[c]{@{}l@{}}\rev{Evaluating prototypes}\\ \rev{with user-controlled} \\ \rev{workloads}\end{tabular} & \begin{tabular}[c]{@{}l@{}}\rev{Highly realistic workloads}\\ \rev{for more accurate} \\ \rev{benchmarking}\end{tabular} \\ \hline
\end{tabular}
}
\vspace{-3ex}
\end{table}

\vspace{-.5ex}
\section{Workload Synthesis in \systemname{}}
\label{sec:redbench}

Since workload traces only contain structural and runtime query metrics and not the exact SQL query, there is a large space of possible query candidates that could all map to the given statistics. 
\systemname{}'s matching-based and generation-based synthesis approaches represent two distinct methods for constraining the search space of candidate queries.
In the following sections, we will first outline the general synthesis flow and then provide more detailed information about each approach.

\begin{figure*}[t]
    \centering
    \vspace{-1ex}
    \includegraphics[width=\linewidth]{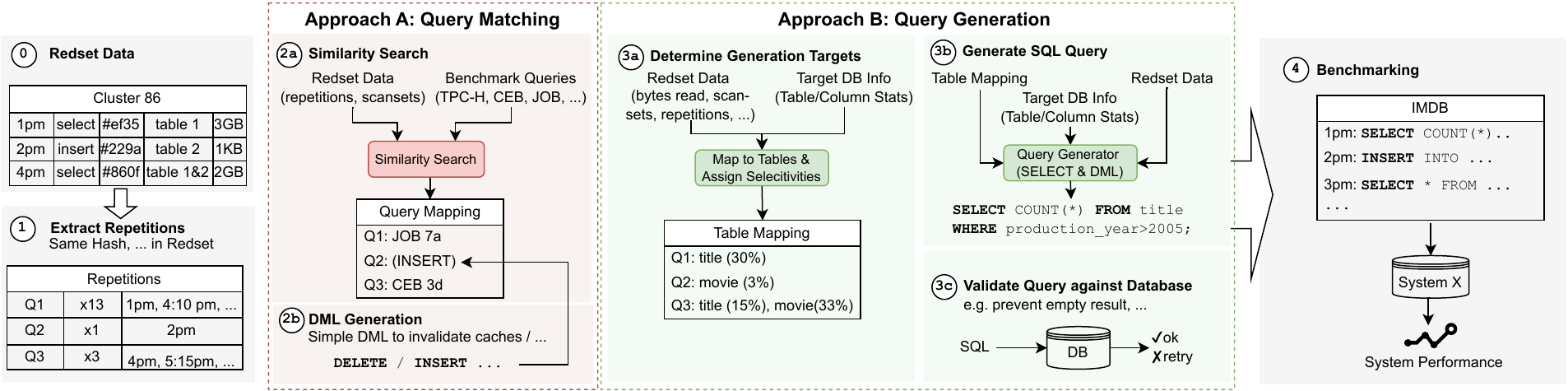}
    \vspace{-5ex}
    \caption{\systemname{} transforms real workload traces \tcircled{0} into executable benchmark workloads. It extracts query repetitions, temporal patterns, and access characteristics from Redset \tcircled{1} and reproduces them by either mapping to existing benchmark queries \tcircled{2} or generating new SQL \tcircled{3}, enabling realistic, workload-driven evaluation of DBMS optimizations \tcircled{4}.}
    \label{fig:architecture}
    \vspace{-2ex}
\end{figure*}

\vspace{-.5ex}
\subsection{Synthesis Flow}
We introduce our approach in  \Cref{fig:architecture}: Workload synthesis is based on a workload trace, such as Redset \circled{0}.
\systemname{} then analyzes the workload trace to infer features like query frequencies, complexities, scansets, repetitions, and temporal locality \circled{1}.

\noindent\textbf{Feature extraction.}
While certain features, such as workload drifts or the sequence of read and write queries, are explicitly encoded in the workload trace \rev{via} timestamps, query repetitions are only implicitly encoded.
Hence, \systemname{} includes a dedicated step \circled{1} to extract \rev{these} implicitly encoded features.
\systemname{} identifies query repetitions by combining the query hash with structural properties, such as the scanset. 
Identifying different types of repetitions, as explained earlier, is crucial for benchmarking and comparing workload-driven optimizations, such as result caching (full query repetitions) versus sub-result caching (scanset repetitions).

\noindent\textbf{Query synthesis.}
The query synthesis step aims to produce a query stream that statistically and temporally resembles the original trace.
It requires both the trace and the dataset, along with its schema as input.
Depending on the synthesis approach, additional inputs may be required, such as a query pool for the matching-based approach.
However, the actual process of query generation is encapsulated in each synthesis approach (\circled{2}~/~\circled{3}), making \systemname{} extensible to other synthesis approaches in the future.

\vspace{-.5ex}
\subsection{Matching-Based Synthesis}
\label{subsec:matching}
The matching-based approach builds on the previous Redbench version~\cite{DBLP:journals/corr/abs-2506-12488}.
\rev{The key idea of this synthesis mode is to map queries in the workload trace to well-designed queries from an existing query pool, such as TPC-H/DS (or, optionally, LLM-generated query pools such as SQLStorm).
In our evaluation, we use human-written queries (JOB and CEB templates) only.}
Hence, it does not generate SQL queries from scratch, but reuses existing queries and orders them to match the workload properties of the original trace. 

\noindent\textbf{Inputs.}
To construct the desired query stream \systemname{} requires two main inputs: (1) a timestamped workload trace from the Redset and (2) a support benchmark such as TPC-DS or JOB that provides a pool of real \emph{query templates}.
We use the support benchmark queries as templates to instantiate different queries with the same query shape but different statistical properties (e.g., bytes scanned).

\noindent\textbf{Synthesis pipeline.}
\systemname{}'s matching-based synthesis uses a deterministic approach to produce a synthetic query timeline that maintains the original arrival order and query mix. 
\rev{Thus}, given the same trace, support benchmark, and seed, \systemname{} will always produce the same synthetic workload.
The generation process shown in \Cref{fig:architecture} (Approach A - red) follows the two basic steps:

\begin{enumerate}
    \setlength\itemindent{0pt}
    \setlength\leftskip{-15pt}
    \item \textbf{Trace processing:} \systemname{} processes the workload trace sequentially to preserve the temporal order and query mix. For each query, it extracts its key statistical metrics, such as the join count or scanset, and determines a suitable mapping.
    \item \textbf{Query mapping (Similarity Search):} For each query, \systemname{} checks if the query has already been encountered or a new query mapping must be created.
\end{enumerate}


\noindent\textbf{General matching strategy.}
The \systemname{} matching strategy \rev{uses} the query hash, scan set, and join count from the workload trace to determine which benchmark query should be mapped to a given trace query.
The main goal is to preserve the \rev{relative number of joins}. 
Specifically, the number of joins in each query is normalized during mapping, i.e., the minimum and maximum number of joins in the Redset workload are linearly mapped to the minimum and maximum number of joins in the support benchmark.

\noindent\textbf{Generating \select{} SQL \circled{2a}.}
To capture repetitions in the workload, \systemname{} keeps track of all synthesized queries it generates. 
\rev{If two \select{} queries from the workload trace share the same hash and constitute full query repetitions, they are mapped to the same \rev{previously generated} benchmark query instance.
\rev{If} the query hashes differ but share the same scanset, the queries \rev{are} mapped to the same query template, and a new query \rev{is} generated.
If no matching candidate could be found, \systemname{} employs a fallback strategy:
If at least one of the closest benchmark templates has unused instantiations, \rev{it} randomly picks one of those.
Otherwise, \rev{it randomly reuses} one of the generated queries from the closest template.}

\noindent\textbf{Generating DML queries \circled{2b}.}
As mentioned earlier, since existing benchmarks mainly focus on read-heavy queries, the matching-based mode \rev{uses} a basic approach \rev{to} generate DML queries.
The primary goal of this generation is to capture read-write dependencies that can lead to, for example, invalidations in caching optimizations.
\rev{Since Redset does not report write statistics (e.g. affected row counts, write volume)}, \systemname{} aligns simple write operations with the corresponding tables of update, insert, or delete operations, acting as invalidators for cached results for subsequent reads.

\vspace{-.5ex}
\subsection{Generation-Based Synthesis}
\label{subsec:generation}
While matching-based synthesis captures many real-world workload properties, it is limited by the support benchmark's fixed query set.
To address this, \systemname{} provides a \emph{generation-based synthesis} mode as a key contribution, creating novel SQL queries from scratch.
This enables greater flexibility in following target workload characteristics.
Beyond read-write patterns and repetitions/arrival times covered by matching, it also models query complexity (num-joins, bytes-read), join paths, and write-heavy queries.


\noindent\textbf{Inputs.}
Generation-based synthesis requires two inputs: 
(1) a workload trace (e.g., from Redset) providing statistics such as repetitions, query types, bytes-read, and query metadata, and
(2) a support database (schema and data) serving as the execution environment for the generated SQL queries.
This can be a public or private dataset.

\noindent\textbf{Synthesis pipeline.}
The generation process shown in \Cref{fig:architecture} (Approach B - green) follows several steps:
\begin{enumerate}
    \vspace{-.7ex}
    \setlength\itemindent{0pt}
    \setlength\leftskip{-15pt}
    \item \textbf{Schema analysis}\rev{~\circled{3a}}: The target schema and data are analyzed to derive a join-graph and table statistics \rev{including the number of rows, data-types of the columns, and value distributions/histogram information about the columns.}
    \item \textbf{Trace property extraction}\rev{~\circled{3a}}:
    From the workload trace, \systemname{} extracts proxy features such as table access patterns (scansets), \rev{query repetitions,} and bytes-read per scanset.
    Aggregating bytes-read across queries yields two key features:
    \rev{(i) the scan-size/bytes-scanned distribution of each table derived from queries accessing the same tables, and
    (ii) an estimated maximum scan size per table, based on the largest observed scan of this table in the Redset trace (derived from single-table queries). 
    This maximum scan size reflects not the true table size, but the largest scan observed in practice.
    The maximum seen scan size will later be used to map to tables in the support benchmark with a similar size, and the bytes-read distribution to determine filter selectivities as explained in the following.}
    \item \textbf{\rev{Size-aware} table mapping}\rev{~\circled{3a}}:
    Tables in the trace are mapped to the tables in the user schema by size; i.e., tables for which we have observed the largest scan sizes \rev{(extracted previously)} are mapped to the largest tables in \rev{the users} schema.
    This size-aware mapping ensures that heavily scanned Redset tables correspond to large physical tables, thereby preserving realistic workload intensity and keeping I/O similar.
    Moreover, it improves predicate generation by leveraging richer data distributions of large tables to generate filters with realistic selectivities (cf. (5)).
    \rev{We assign target selectivities for the filters of each table to match the reported bytes-scanned of the Redset trace.
    Since the Redset trace does not exhibit per-table scan-sizes for multi-table queries, we assign selectivities for each table referenced in the Redset-trace query by sampling on the bytes-read distribution observed from single table queries described in (2).}
    \item \textbf{Join-path generation}\rev{~\circled{3b}}:
    \rev{Starting from the mapped support-database table of the (assumed) largest table in the Redset query, \systemname{} performs random walks over the join-graph of the users dataset schema until the Redset trace's number of joins is met}. 
    \rev{Since the exact join-path and schema of the Redset trace can only be speculated, join-path generation can only be approximate. 
    This mapping might differ from the original (unknown) join-path depending on the the number of tables and structure (e.g., star schema) of the (unknown) Redset db schema and the user's support database 
    }
    \item \textbf{Predicate assignment and selectivity control}\rev{~\circled{3b}}:
    \rev{Having assigned a target selectivity for each query-table, and mapped the Redset-tables involved in the query to tables in the users schema, filter-columns are randomly chosen for each of the mapped tables.
    Utilizing the previously extracted histogram information of the columns from (1), filter-literals that match the assigned target selectivities are chosen.
    To closely match the assigned target selectivities and, through that, match the overall bytes-read, point-lookup, and range predicates on columns of arbitrary data types are sampled, since their selectivities can be tuned with reasonable effort using the collected histogram information and by refraining from executing the queries.
    }
    \vspace{-1ex}
\end{enumerate}

\noindent\textbf{Generating \select{} SQL.}
Approximately 50\% of all queries in Redset are \SELECT{} queries.
These are referred to as read-only queries and play an important role in workload simulation, particularly because they are typically cacheable.
Depending on the specific customer, the fraction of read-only queries can vary from none to all, making the accurate generation of these queries essential for capturing realistic usage patterns.
To produce such queries, the table mapping and selectivity assignment from \circled{3a}, together with statistics about the target database and additional information from Redset (num aggregates), are used to produce a matching SQL query \circled{3b}.
Lastly, the generated queries are validated against the support database to, e.g., avoid empty results \circled{3c}\rev{, however, not to tune their performance or bytes-read characteristics}.
This step is necessary because only per-table statistics are used and predicates are chosen per table, so empty results cannot be ruled out by design. Validation ensures that smart work-skipping by the DBMS does not distort the evaluation.
    
\noindent\textbf{Generating DML queries.}
About half of all Redset queries are DMLs, so matching their complexity is critical for realistic workload distributions and read-write heaviness.
The largest DML type, with approximately 20\% of Redset queries, are \INSERT{} statements, most of which are \INSERT{}~/~\SELECT{} queries with a read subquery.
Generating meaningful data for such queries synthetically is challenging, as they resemble materialization steps that require a logical relationship between source (read) and target (write) tables.
To capture this, \systemname{} applies a schema preprocessing step that creates a \emph{staging table} for each table.
These are copies of the original tables, \rev{carrying forward} primary and foreign keys so each row can be inserted into the original table without violating PK/FK constraints (also enabling non-empty join results).
Non-PK/FK columns with unique values receive suffixes so that, after insertion, \rev{column uniqueness constraints} remain intact, preserving \rev{data distributions similar to the original base tables}.
These staging and original tables then serve as both sources and destinations for \INSERT{} operations, ensuring schema compatibility and scalable query generation.
Since Redset does not contain information about the number of rows updated by DMLs, \systemname{} randomly selects a number of rows to modify, ensuring the overall data distribution remains stable and roughly matches the Redset \rev{read-write ratio}.

Notably, in Redset, \INSERT~/~\SELECT{} queries often exhibit high repetition, indicated by identical scansets, query hashes, and read volume.
Hence, the same read-table mapping used for \SELECT{} queries is applied here as well, allowing \systemname{} to reproduce and analyze caching effects on the read portions of write queries (as long as their read-tables have not changed), \rev{similar} to read-only workloads.

\noindent\textbf{Special handling of \UPDATE{} and \DELETE{}.}
For \UPDATE{} and \DELETE{} queries, \systemname{} generates lightweight operations that modify a small, randomly selected subset of rows to roughly match the Redset\rev{'s write load} while preserving data distribution stability. 
Hence, they are structurally similar to \SELECT{} queries but differ in that they write or delete \rev{rows in} a table. 
For \UPDATE{}s, the generator selects a non-primary, non-foreign key column of the write-table \rev{to} update with new values. 
The new values are sampled randomly based on column statistics.
Other operations, such as \COPY{} or \ANALYZE{}, are omitted because they are not consistently supported across systems (or differ significantly in purpose between the systems) and add limited value for evaluating workload-driven optimizations.

\section{Evaluation using \systemname{}}

We now evaluate \systemname{} on \emph{production-grade cloud data warehouses} (Systems~A--D) with various workload-driven optimizations. 
Our experiments show why a realistic, trace-shaped benchmark is necessary and what traditional benchmarks cannot reveal. 
\vspace{-1ex}
\subsection{Goals of Evaluation}
Our evaluation answers two central questions: (1) How well does \systemname{} reproduce the statistical, temporal, and repetition properties of real-world traces? 
(2) Why does this matter—i.e., does \systemname{} expose new optimization effects (e.g., caching, reuse) that traditional benchmarks miss? 
Upfront, we show that traditional benchmark baselines suggest negligible caching benefits, while \systemname{} reveals substantial speedups on the \emph{same systems}, underscoring the necessity of cloud-like, trace-shaped workloads.

\vspace{-1ex}
\subsection{Experimental Setup}
\label{subsec:exp_setup}
\begin{figure*}
    \vspace{-1ex}
    \centering
    \includegraphics[width=\linewidth]{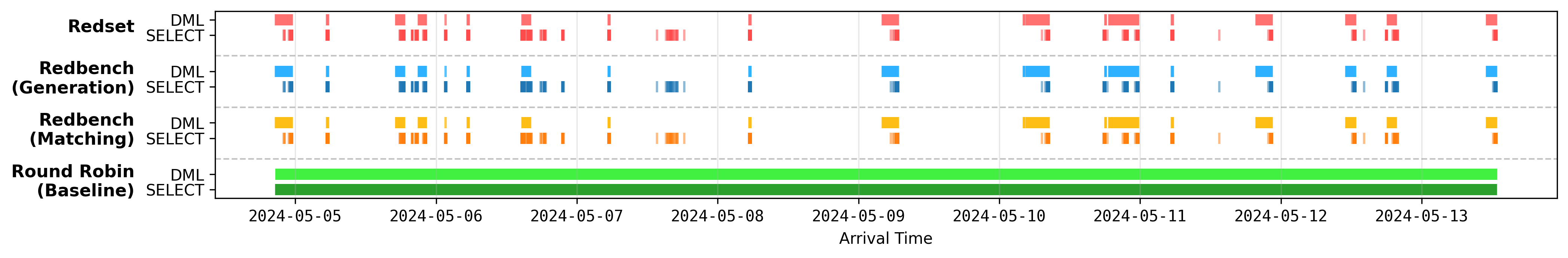}
    \vspace{-6ex}
    \caption{\systemname{} faithfully preserves the temporal arrival patterns from Redset, including distinct bursts of activity followed by idle periods. Both generation-based and matching-based approaches maintain exact arrival times, ensuring realistic temporal dynamics for evaluating workload-driven optimizations.}
    \label{fig:arrival_pattern}
    \vspace{-1ex}
\end{figure*}

To evaluate \systemname, we \rev{use} the publicly available Redset~\cite{DBLP:journals/pvldb/RenenHPVDNLSKK24} and the real-world IMDb dataset for workload synthesis.

\noindent\textbf{Cluster selection.}
\label{par:cluster_selection}
\rev{From the Redset provisioned dataset, we select a representative cluster and database\footnote{\textit{Dataset}: Redset-Provisioned, \textit{Cluster}: 186, \textit{Database}: 8, \textit{Timeframe}: May 2024} \rev{with} a balanced mix of query types (\SELECT, \INSERT, \UPDATE, \DELETE) and repetition patterns close to the averages reported in \cite{DBLP:journals/pvldb/RenenHPVDNLSKK24}.}
The trace contains 36k queries: 25k \SELECT{} queries (6.4k unique) and 11k diverse DML queries (\INSERT, \UPDATE, \DELETE).
The query repetition rate (QRR) is representative at 73\% as shown in \Cref{tab:gen_wl_stats}.

\noindent\textbf{Support database.}
In our \rev{e}valuation, we generate workloads on the real-world IMDb dataset to enable realistic data distributions and string-heavy queries --- a key characteristic observed in Redset.
The IMDb schema allows us to build on sophisticated support benchmarks such as JOB~\cite{DBLP:journals/pvldb/LeisGMBK015} and CEB~\cite{DBLP:journals/pvldb/NegiMKMTKA21}.
However, \systemname{} can also use other databases as support databases for the synthesis.

\noindent\textbf{Evaluation on four cloud systems.}
We evaluate \systemname{} on four commercial cloud data warehouse systems (Systems A--D) that implement various workload-driven optimizations such as result caching, predicate caching, and scan-set pruning.
They represent diverse architectures and optimization strategies commonly employed in modern cloud data warehouses, ranging from industry-leading cloud-native systems to aspiring newcomers.
The generated queries are executed sequentially with compressed idle gaps to isolate the impact of workload-driven optimizations and avoid confounding effects from concurrency control or resource contention that would arise in real-time execution.
While evaluating these concurrency aspects would be valuable and is possible with \systemname{}, it requires access to internal DBMS statistics that are typically not publicly exposed and are highly system-specific.


\noindent\textbf{Baseline benchmark.}
\rev{As discussed in \Cref{subsec:related_wl_synth}, with the exception of PBench, other existing benchmarks were not designed to reproduce the characteristics of real workload traces.
However, due to reproducibility issues with PBench and despite greater patching efforts, we were not able to use it as a baseline.
Hence, to evaluate the effectiveness of \systemname{}'s workload generation approaches, we compare against a baseline based on traditional benchmarks, called the \emph{Round-Robin baseline}, which selects queries solely based on read/write proportions.}
Instead of matching or generating queries based on the Redset trace properties, this naive baseline cycles through queries from the support benchmark (i.e., JOB \& CEB)  in a round-robin \rev{order}.
We select as many distinct \SELECT{} queries as there are unique read queries in the Redset trace.
To approximate the temporal query mix, the baseline follows a fixed read/write pattern based on the overall ratio observed in Redset (e.g., a 2:1 read/write ratio becomes $r,r,w,r,r,w,\ldots$).
\rev{Since JOB and CEB contain only read queries, we inject minimal DML invalidators (lightweight single-row delete/insert statements, analogous to our matching-based synthesis)} to simulate cache invalidations.
Arrival times are uniformly interpolated between the start and end of the original trace, resulting in evenly spaced queries.

\begin{table}
\centering
\caption{Query Repetition Rate (QRR) and Scan-set Repetition Rate (SRR) denote the fraction of \SELECT{} queries that repeat previous queries (QRR) or scan-sets (SRR).
QRR$_{\text{DML}}$ and SRR$_{\text{DML}}$ indicate DML-aware repetition rates.
Generation-based synthesis closely matches the original Redset trace, while matching-based synthesis produces higher repetition due to reusing a limited query pool.
The baseline fails to capture realistic repetition patterns.
}
\label{tab:gen_wl_stats}
\vspace{-2ex}
\small
\begin{tabular}{@{}lcccc@{}}
\toprule
\textbf{Approach} & \textbf{QRR$_{\text{DML}}$} & \textbf{QRR} & \textbf{SRR$_{\text{DML}}$} & \textbf{SRR} \\ 
\midrule
Redset (Original) & 0.72 & 0.73 & 0.86 & 0.87 \\
\systemname{} (Generation) & 0.73 & 0.79 & 0.95 & 0.98 \\
\systemname{} (Matching) & 0.97 & 0.98 & 0.99 & 0.99 \\
Round-Robin Baseline & 0.00 & 0.74 & 0.36 & 0.99 \\
\bottomrule
\end{tabular}
\vspace{-4ex}
\end{table}

\vspace{-.5ex}
\subsection{Faithfulness of \systemname{}}
\label{subsec:synthesis_quality}

To enable realistic evaluation of workload-driven optimizations, a benchmark must faithfully reproduce the key characteristics of production workloads.
We first demonstrate that \systemname{} achieves this goal by closely matching the statistical and temporal properties across four critical dimensions: (1) temporal dynamics and query type distributions, (2) query repetition patterns, (3) system load characteristics, and (4) DML-aware invalidation effects.
The round-robin baseline, despite matching overall query counts and read/write ratios, fails to capture these production workload properties, confirming the necessity of \systemname{}'s trace-driven approach.

\noindent\textbf{Temporal patterns and query type distributions.}
Production workloads exhibit complex temporal patterns, including bursts, idle \rev{periods}, and time-varying read/write mixtures, which are essential for evaluating adaptive database optimizations.
\systemname{} preserves these patterns by retaining the exact arrival times from the Redset trace.
\Cref{fig:arrival_pattern} demonstrates that \systemname{} faithfully reproduces the bursty activity patterns observed in Redset, with distinct periods of high query load followed by idle intervals.

Furthermore, \systemname{} preserves query type distributions exactly: each generated or matched query maintains the same type (\SELECT, \INSERT, \UPDATE, \DELETE) as its corresponding trace entry.
This precise matching of read/write interdependencies is critical: DML operations invalidate caches at the same temporal positions as in production, triggering realistic cache invalidation cascades and exposing the effectiveness of optimizations under authentic write pressure.
For the matching-based approach, however, DML queries \rev{may} be missing in its query pool. 
Therefore, minimal DML statements are generated by \systemname{} and added to the pool.
This ensures, for both approaches (generation and matching), temporal alignment of cache invalidations with the original trace.
In contrast, the round-robin baseline distributes queries uniformly over time with a fixed read/write interleaving pattern (e.g., $r,r,w,r,r,w,\ldots$), missing the bursty and irregular temporal dynamics of real workloads.
\begin{table}[t]
\caption{\rev{Data type distribution of filter columns. TPC-DS underrepresents string-typed filters (22\%) compared to Snowflake's real-world workload statistics (58\%)~\cite{DBLP:journals/pvldb/SzlangBCDFHOOM25}. \systemname{} yields a more realistic string-heaviness(67\%/75\%). It only uses numeric and string types because the underlying IMDb dataset does not contain other data types.}}

\label{tab:dtypes}
\vspace{-2ex}
\small
\changedimage{
\begin{tabular}{lccc}
\toprule
\textbf{Benchmark} & \textbf{Text} & \textbf{Numeric} & \textbf{Other} \\
\midrule
Snowflake Cloud Statistics~\cite{DBLP:journals/pvldb/SzlangBCDFHOOM25} & \textbf{58\%} & 25\% & 17\% \\
TPC-DS~\cite{DBLP:journals/pvldb/SzlangBCDFHOOM25} & \textbf{22\%} & 71\% & 7\% \\
\systemname{} (matching) \textit{w/ IMDb} & \textbf{67\%} & 33\% & n/a \\
\systemname{} (generation) \textit{w/ IMDb} & \textbf{75\%} & 25\% & n/a \\
\bottomrule
\end{tabular}
}
\vspace{-3ex}    
\end{table}
\rev{Further, \Cref{tab:dtypes} shows that a realistic string-heavy workload, similar to production workloads described by \cite{DBLP:journals/pvldb/SzlangBCDFHOOM25} is produced.
\systemname{} comes with a ratio of 67\%/75\% (matching/generation) filters operating on string-columns on the IMDb dataset.
This closely matches and even exceeds statistics of production traces that come with 58\%\footnote{\rev{Statistics from Snowflake \cite{DBLP:journals/pvldb/SzlangBCDFHOOM25}. Redset\cite{DBLP:journals/pvldb/RenenHPVDNLSKK24} does not report data type statistics.}} of filters operating on string columns\cite{DBLP:journals/pvldb/SzlangBCDFHOOM25}.
Though, \systemname{}'s string heaviness is influenced by the presence of string-columns / string-heavy queries in the support database / workload. \\}
\begin{figure}
    \vspace{-1ex}
    \centering
    \changedimage{\includegraphics[width=\linewidth]{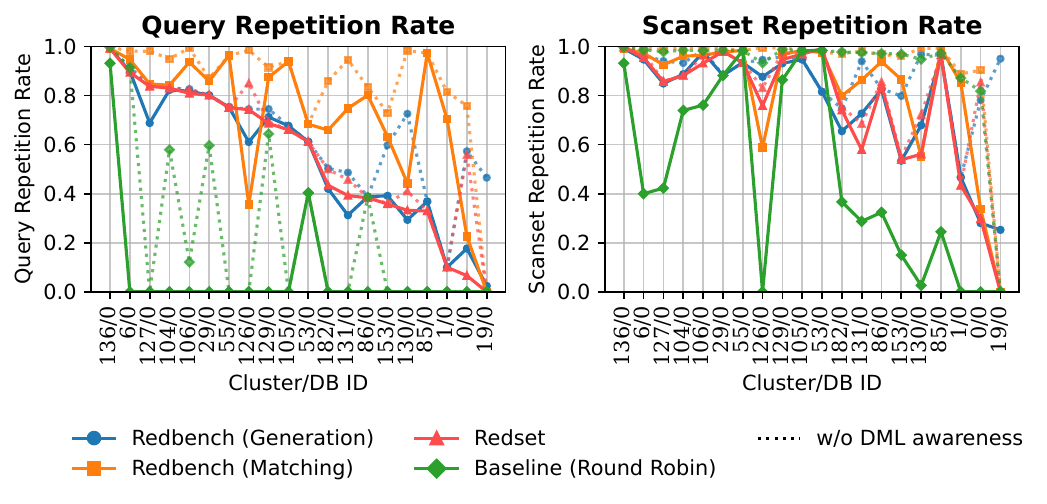}}
    \vspace{-5.5ex}
    \caption{Comparison of query repetition rate (QRR) and scan-set repetition rate (SRR) between \systemname{} and Redset. 
    DML-aware metrics consider cache invalidations from interleaved writes on the same tables.
    Generation-based synthesis closely matches Redset's repetition characteristics, while matching-based exhibits higher repetition due to the limited query pool. 
    The baseline fails to replicate realistic repetition patterns.
    }
    \label{fig:repetition_rates}
    \vspace{-4ex}
\end{figure}

\noindent\textbf{Query repetition patterns.}
\rev{Repeating query patterns are fundamental characteristics of production workloads, directly determining the effectiveness of result caching, intermediate materialization, and incremental query processing.
We measure two key metrics: query repetition rate (QRR), indicating exact query re-execution, and scan-set repetition rate (SRR), capturing repeated table access patterns even for structurally different queries.
We further consider that DMLs invalidate cached results if based on an updated table.
QRR/SRR$_{\text{DML}}$ describe \textit{DML-aware} repetitions where no invalidating \INSERT{}, \UPDATE{}, or \DELETE{} occurred in between.}

\Cref{tab:gen_wl_stats} shows that \systemname{}'s generation-based mode closely matches Redset's repetition characteristics (QRR$_{\text{DML}}$: 0.73 vs. 0.72) on the Redset cluster described in \Cref{par:cluster_selection}, demonstrating realistic query diversity despite mapping to a different schema.
The matching-based approach yields higher repetition rates (QRR: 0.98) because it operates on a limited query pool, \rev{forcing reuse of} the same SQL query for different Redset queries.
Hence, this represents an upper bound on caching potential.
Both approaches substantially outperform the round-robin baseline, achieving 0\% DML-aware QRR despite identical query counts.
Its uniform query distribution prevents realistic repetition patterns.

\rev{Having a single Redset cluster analyzed, we \rev{now} consider the entire cluster pool in the Redset dataset.}
\Cref{fig:repetition_rates} shows repetition rates for 20 randomly sampled clusters.
Across all clusters, \systemname{}'s generation-based approach closely matches Redset's repetition patterns, whereas the matching-based approach yields higher repetition due to its limited query pool.
The round-robin baseline fails to replicate realistic repetition patterns and mostly does not follow Redset; in contrast, it exhibits no repetitions, although Redset with the same number of unique queries shows significant repetition rates.

Notably, \systemname{} achieves high fidelity to the trace despite a significant challenge: the IMDb support schema contains fewer tables than the original Redset database, forcing multiple Redset tables to map onto a single IMDb table.
This schema mismatch \rev{complicates maintaining} consistent repetition rates, yet both synthesis approaches successfully reproduce realistic patterns.
This robustness is critical for practical adoption, as original production schemas of Redset or other workload traces are not accessible \rev{and} users must rely on public datasets.

\rev{Furthermore, as shown in \Cref{fig:repetition_rates}, DML operations substantially affect effective repetition rates.
DML-aware rates are consistently lower than their non-DML-aware counterparts, quantifying how \rev{interleaved} writes reduce opportunities for repetition-based wl-driven optimizations such as result caching.}
\rev{Depending on the workload, this effect can vary in magnitude. 
For example, heavy ETL or ingestion phases \rev{near} frequently repeated \rev{queries} invalidate cached results more often, leading to a larger drop in DML-aware repetition rates.
This \rev{underscores} the importance of DML-awareness in workload synthesis for realistic caching evaluation.}

\begin{figure}
    \vspace{-1ex}
    \centering
    \includegraphics[width=.85\linewidth]{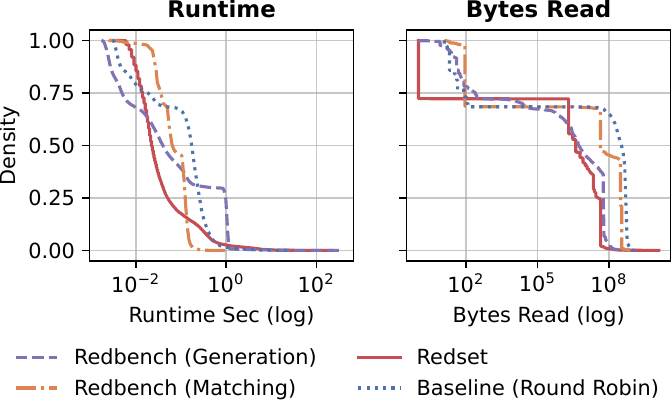}
    \vspace{-3ex}
    \caption{CDF of runtime and bytes-read (log-scale) for Redset, \systemname{} (generation \& matching) and the round-robin baseline. 
    Both synthesis methods closely match the overall load distribution of Redset, with generation-based achieving a tighter fit in the tail through predicate adaptation.
    The baseline shows fundamentally different load characteristics.}
    \label{fig:similarity_cdf}
    \vspace{-3ex}
\end{figure}

\noindent\textbf{System load characteristics.}
\rev{Beyond temporal and repetition patterns, realistic workload synthesis must also match system load distributions (\rev{notably} bytes-read and query runtimes) to evaluate \rev{optimizations under realistic} resource pressure.
\Cref{fig:similarity_cdf} compares runtime and bytes-read \rev{distributions} between Redset and the round-robin baseline.
The generation-based approach \rev{closely aligns} with Redset's load distribution by explicitly steering filter selectivity \rev{toward} target bytes-read values.
\rev{Its} predicate adaptation produces queries with similar I/O footprints to the original trace, resulting in comparable runtimes, especially in the long tail, despite not directly targeting execution time.
The matching-based approach selects queries solely \rev{by} structural complexity (e.g., number of joins) \rev{and therefore aligns less well with Redset, as it lacks} explicit bytes-read steering.
This is \rev{particularly} visible in the distribution tail, where complex queries \rev{exhibit higher} variability in resource consumption that the matching-based approach \rev{does not capture}.}

Both \systemname{} approaches outperform the round-robin baseline, which shows different load characteristics \rev{since} it does not match any Redset workload properties beyond overall query counts.
The close \rev{alignment} of runtime and bytes-read distributions confirms that \systemname{} generates workloads with realistic system stress patterns, essential for evaluating performance-oriented optimizations.

\begin{figure}
    \centering
    \vspace{-.5ex}
    \changedimage{\includegraphics[width=.9\linewidth]{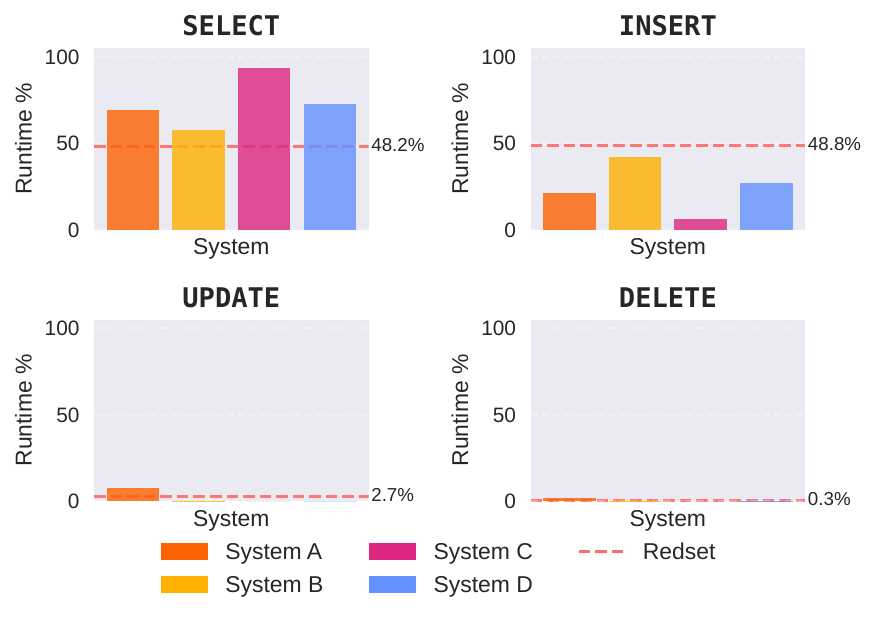}}
    \vspace{-2.5ex}
    \caption{Runtime distribution across query types (\SELECT, \INSERT, \UPDATE, \DELETE) for Redset \rev{(numbers computed on the trace of ~\cite{DBLP:journals/pvldb/RenenHPVDNLSKK24} shown with ``\textcolor{red}{\texttt{-}\texttt{-}\texttt{-}}'')} and \systemname{} (generation-based) on four cloud database systems (no caching active).
    System-specific deviations reveal optimization differences \rev{on Redbench}: System~C prioritizes fast \INSERT{}s, System~B exhibits fast \SELECT{}s, and System~A shows \DELETE{} slowdowns.
    }
    \label{fig:query_type_runtimes}
    \vspace{-4ex}
\end{figure}

\noindent\textbf{System comparison.}
\Cref{fig:query_type_runtimes} further validates system load fidelity by comparing runtime distributions across query types (\SELECT{}, \INSERT{}, \UPDATE{}, \DELETE{}) between Redset and \systemname{}'s generation-based approach on four cloud data warehouses.
The matching-based and baseline approaches are excluded, as they only include simple DML queries \rev{that} do not \rev{reflect} realistic DMLs.
\systemname{} (generation-based) achieves alignment with Redset's query-type runtime distribution. 
This confirms that synthesized workloads impose realistic resource demands for both read and write operations.
However, depending on \rev{how} the underlying system \rev{is optimized for} DMLs, the writes will amount to varying magnitudes in the overall workload runtime.
Finally, \Cref{fig:query_type_runtimes} reveals another important aspect:
\systemname{} \rev{enables comparison of} system-specific differences \rev{stemming from distinct} optimization strategies.
For instance, System~C achieves fast \INSERT{} performance but exhibits proportionally higher SELECT runtime, suggesting write-optimized storage structures.
System~B, in contrast, trades ingestion \rev{for} read-performance, \rev{reflected in} elevated \INSERT{} costs, indicating potential compromises in write\rev{-}path optimizations.
System~A demonstrates \DELETE{} slowdowns, possibly due to complex index maintenance or integrity checking.
These deviations highlight an important benefit: while \systemname{} faithfully reproduces workload characteristics, cross-system comparison exposes architectural trade-offs and optimization priorities that remain hidden in single-system evaluations.

\noindent\textbf{Summary.}
Across all four evaluation dimensions --- temporal dynamics, repetition patterns, system load, and DML-aware invalidation effects --- \systemname{} demonstrates high fidelity to production workloads.
The generation-based approach provides the closest match to Redset's statistical properties while preserving query diversity, making it ideal for trace-faithful benchmarking.
The matching-based approach provides an upper bound on the potential of repetition-driven optimization, making it useful for stress-testing caching mechanisms.
Both approaches dramatically outperform the round-robin baseline, which fails to capture realistic workload properties despite matching aggregate statistics.

\vspace{-.5ex}
\subsection{Effects of Caching in \systemname{}}
\label{subsec:eval_caching}

\begin{figure}
    \vspace{-.5ex}
    \centering
    \changedimage{\includegraphics[width=\linewidth]{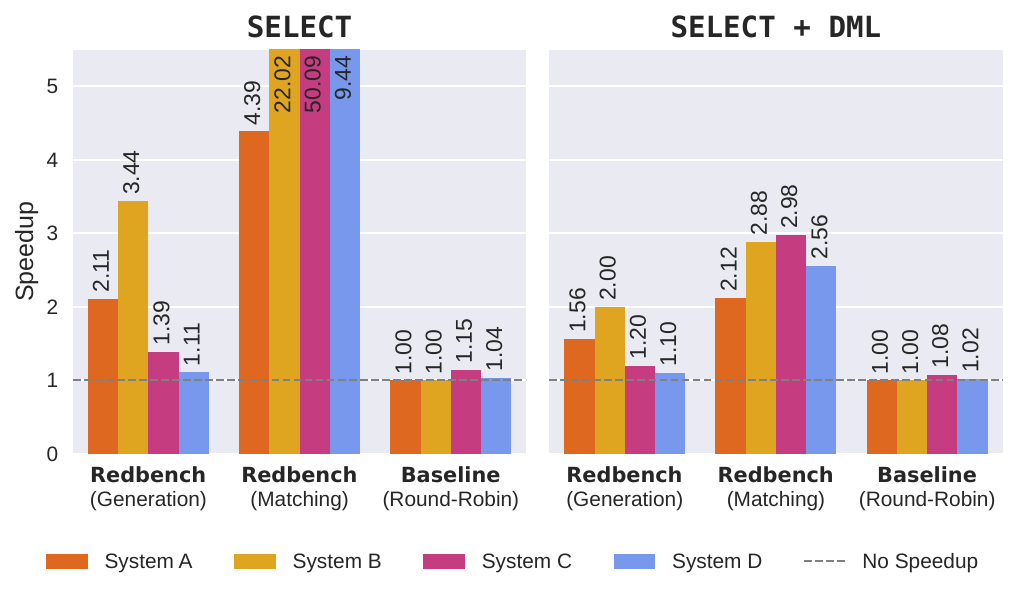}}
    \vspace{-5ex}
    \caption{Overall caching speedups across systems for \systemname{} (generation-based, matching-based) and a traditional-benchmark baseline (JOB \& CEB).
    Traditional benchmarks show no speedup, (misleadingly) suggesting that caching is ineffective.
    On the \textit{same systems}, \systemname{} workloads reveal substantial caching potential: generation-based yields moderate, trace-faithful improvements (1.10--2.00$\times$), while matching-based demonstrates upper-bound caching effects (2.12--2.98$\times$) due to higher repetition density.
    }
    \label{fig:cache_speedups}
    \vspace{-4ex}
\end{figure}

In the next set of experiments, we \rev{show} that \systemname{}, which generates production-like workloads, can reveal important effects of system optimizations that traditional benchmarks cannot.
In particular, to demonstrate this ability, we focus on how \systemname{} reveal\rev{s} the effectiveness of various forms of caching recently introduced in cloud data warehouses.
However, we believe that \systemname{} can \rev{also} provide insights into many other optimizations and guide research on which optimizations are effective and which are not.

\noindent\textbf{Traditional benchmarks fail to expose caching opportunities.}
\Cref{fig:cache_speedups} reveals a critical limitation of traditional benchmarks for evaluating the effects of caching: when executed on the round-robin baseline derived from JOB and CEB, all four systems \rev{show} negligible result caching speedups ($\leq$1.08$\times$) --- almost no \rev{benefit from} caching.
This \rev{is} because traditional benchmarks lack realistic query repetition patterns and temporal correlation.
Queries arrive in deterministic round-robin order \rev{without} bursts or idle periods, and DML operations appear uniformly across tables in static patterns.
Based solely on these results, one might conclude that result caching provides minimal benefit and is not worth implementing.

\rev{However, existing benchmarks fail to capture realistic arrival and repetition behavior.
\rev{Unlike} traditional benchmarks, workloads generated by \systemname{} reveal substantial caching opportunities that traditional benchmarks \rev{entirely} miss.
The generation-based approach, which \rev{most closely matches} the original trace's statistical properties (cf. \Cref{subsec:synthesis_quality}), yields speedups due to caching ranging from 1.10$\times$ to 2.00$\times$ across systems.}
These moderate but realistic improvements reflect the repetition rates and read-write interleaving observed in production traces, where caching opportunities exist but are limited by frequent invalidation due to writes.

The matching-based variant of \systemname{} produces even stronger effects (2.12$\times$ to 2.98$\times$ speedups) due to its higher repetition from reusing queries from a limited query pool.
While less representative of typical production workloads, this approach establishes an upper bound on caching potential and is valuable for stress-testing cache implementations.
Critically, both \systemname{} variants (matching and generation) expose caching behavior that traditional benchmarks fail to trigger, demonstrating that realistic workload synthesis is essential for evaluating workload-driven optimizations that would otherwise be \rev{overlooked}.

\begin{figure*}[t!]
    \centering
    \begin{subfigure}[b]{0.33\linewidth}
        \includegraphics[width=\linewidth]{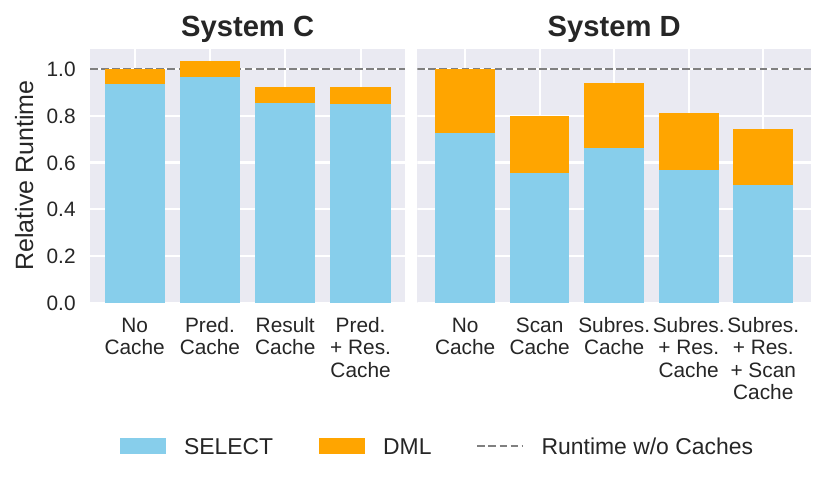}
        \vspace{-4ex}
        \caption{\systemname{} (Generation)}
        \label{fig:cache_drilldown_generation}
    \end{subfigure}
    \hfill
    \begin{subfigure}[b]{0.33\linewidth}
        \includegraphics[width=\linewidth]{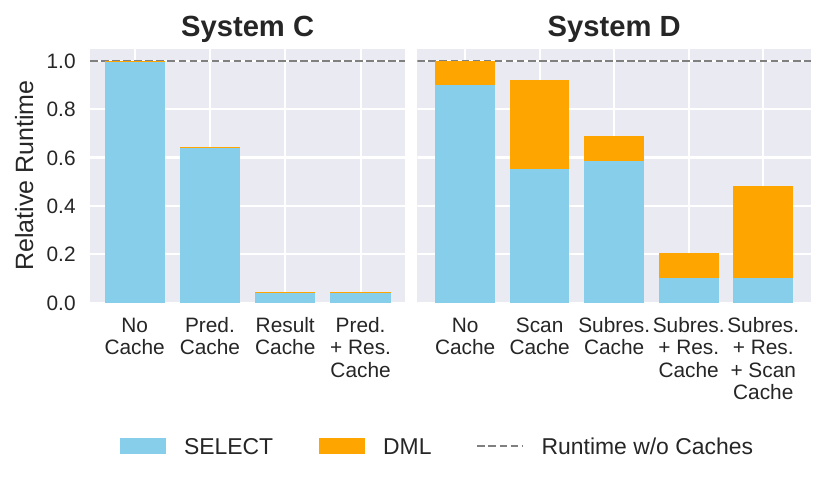}
        \vspace{-4ex}
        \caption{\systemname{} (Matching)}
        \label{fig:cache_drilldown_matching}
    \end{subfigure}
    \hfill
    \begin{subfigure}[b]{0.33\linewidth}
        \includegraphics[width=\linewidth]{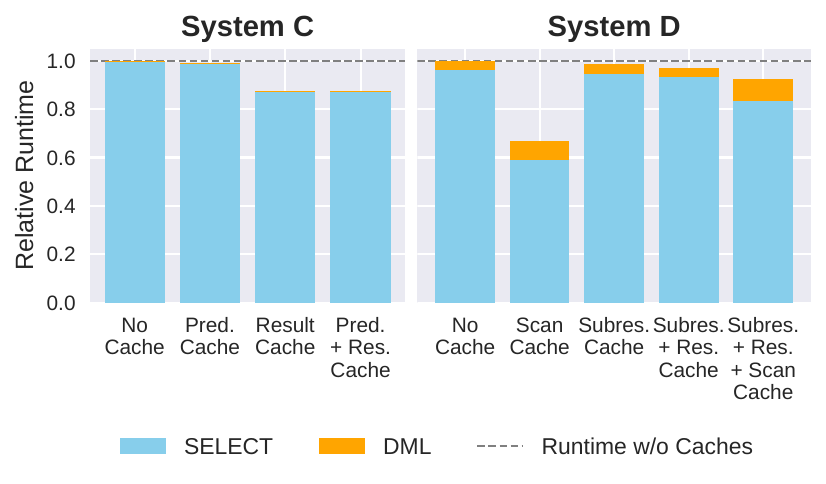}
        \vspace{-4ex}
        \caption{Round-Robin Baseline}
        \label{fig:cache_drilldown_round_robin}
    \end{subfigure}
    \vspace{-4.5ex}
    \caption{Runtime under different caching configurations for generation-based, matching-based, and baseline workloads. 
    Result caching provides clear benefits with minimal overhead across workloads.
    More fine-grained mechanisms (predicate-, subresult-, scan-caches) offer limited additional benefit and can regress performance when frequent DML operations trigger invalidation overhead or maintenance costs.
    The round-robin baseline exhibits different behavior: scan caching shows small speedups due to predictable table access patterns, but result caching remains ineffective because there is no genuine query repetition.
    }
    \label{fig:cache_drilldowns}
    \vspace{-2.5ex}
\end{figure*}

\noindent\textbf{Learnings from caching analysis.}
Next, we reveal more nuanced insights from \systemname{} for caching.
While result caching \rev{shows} clear benefits across all systems, the fine-grained breakdown in \Cref{fig:cache_drilldowns} reveals that different systems employ markedly different caching strategies.
\Cref{fig:cache_drilldowns} compares runtime under different caching configurations for generation-based, matching-based, and baseline workloads of System~C and System~D, showing their impact on \SELECT{} and DML runtime.
These architectural differences expose critical trade-offs that only become apparent under realistic workload conditions.
\rev{We examine System~C and System~D in detail, as they provide more fine-grained caching mechanisms representing contrasting design philosophies.} 
While System~A and System~B only offer result-caching, System~C and System~D \rev{support} predicate, subresult, and scan caching.
Predicate caching stores outcomes of individual predicate evaluations (e.g., \texttt{col>10}) for reuse in later queries, pruning page accesses.
Subresult caching retains intermediate join or aggregation results \rev{for reuse} by subsequent queries.
Scan caching keeps table blocks or partitions in memory to avoid redundant I/O on repeated scans.
We evaluated these configurations by enabling/disabling them individually and measuring their isolated and combined effects on our generated workloads.

\noindent\textbf{System~C: Query-level caching excels under high repetition.}
System~C provides several caching configuration knobs: \textit{result caching} (caching complete query results), \textit{predicate caching} (reusing predicate evaluations across queries), and combinations thereof.
Under the generation-based workload (\Cref{fig:cache_drilldown_generation}), which closely matches production trace characteristics with moderate repetition rates and realistic read-write interleaving, System~C shows mixed results.
\textit{Result caching} yields only modest speedups (\rev{about} 10\% runtime reduction), indicating moderate benefits of result reuse in System~C.
In contrast, \textit{predicate caching} causes performance regressions.
This slowdown can be attributed to DML operations (\INSERT{}, \UPDATE{}, \DELETE{}) that increase in runtime.
\rev{We believe this slowdown stems from checking and invalidating cached predicates, which in System~C appears to have \rev{higher} overhead than checking and invalidating full result-cache entries.
Hence, \rev{unlike} coarse-grained \textit{result caching}, \rev{which provides} speedups with little overhead, fine-grained \textit{predicate caching}, while promising in theory, becomes counterproductive because the cost of maintaining and invalidating cached predicates exceeds their benefit.}
System~C's true strength emerges under the matching-based workload (\Cref{fig:cache_drilldown_matching}), which features substantially higher query repetition due to reuse of a limited query pool.
Here, System~C achieves the highest overall runtime reductions.
The configuration breakdown reveals why System~C excels under high-repetition conditions:
\textit{Predicate caching} contributes 35\% runtime reduction, providing minor but measurable incremental benefits in this read-heavy, high-repetition scenario.
\textit{Result caching} brings runtime down by about 90\%, demonstrating its effectiveness as a coarse-grained optimization with minimal overhead when exact query matches dominate.
With the round-robin baseline (\Cref{fig:cache_drilldown_round_robin}), only minor speedups are observed with result caching and no significant benefit from scan caching, \rev{consistent} with the overall negligible speedups seen previously (\Cref{fig:cache_speedups}), since only few query repetitions occur in this baseline.
Lastly, we see that these different optimizations do not expose overheads on the write-path, with DML query runtime remaining constant across all workloads.

Overall, System~C excels under high query repetition rates, where result caching provides the dominant benefit with minimal overhead.
Predicate caching offers only minor incremental value and\rev{,} in realistic write-interleaved workloads\rev{,} can introduce overhead that outweighs its benefits.
All these optimizations have no effect on write path performance.

\noindent\textbf{System~D: Fine-grained caching for write-interleaved workloads.}
System~D features a sophisticated caching hierarchy with multiple configurations: \textit{scan caching} (block-level), \textit{subresult caching} (intermediate join results), and combinations with \textit{result caching}.

Under the generation-based workload (\Cref{fig:cache_drilldown_generation}), which aligns most closely with production trace characteristics, System~D demonstrates the effectiveness of its fine-grained caching mechanisms. 
First, we must note that this system is less optimized for DML operations, \rev{which} account for 30\% of overall runtime, compared to just 10\% in System~C. 
Considering the caching options, the generation-based approach \rev{shows} a 15\% reduction in total runtime for \textit{scan caching} by reusing recently accessed data blocks, while \textit{subresult caching} alone contributes an 8\% reduction. 
When combined with \textit{result caching}, these mechanisms achieve a maximum runtime reduction of 25\%, highlighting the\rev{ir} complementary nature.

In the matching-based workload, System~D's advantages become even clearer (\Cref{fig:cache_drilldown_matching}), as higher query repetition rates of this synthesis variant lead to runtime reductions of up to 75\%. 
While \textit{scan caching} offers modest speedups, it also increases the proportion of DML runtime from 10\% to 40\%. 
Conversely, \textit{subresult caching} maintains lower DML costs while achieving similar \SELECT{} performance to scan caching. 
\rev{The best results come from combining subresult \rev{and} result caching, \rev{yielding} a 75\% runtime \rev{reduction} without significantly increasing DML costs.}
Under the round-robin baseline (\Cref{fig:cache_drilldown_round_robin}), System~D shows minimal speedups across configurations due to low query repetition. 
While \textit{scan caching} provides slight benefits from predictable access patterns, these gains are artificial and do not reflect realistic workload behavior.

Overall, System~D's fine-grained caching mechanisms enhance performance in write-interleaved workloads by enabling partial result reuse. 
However, they introduce complexity and maintenance overhead, particularly with frequent DML operations that trigger invalidations. 
A careful balance between caching granularity, overhead, and workload characteristics is essential for optimizing performance, as all caching mechanisms contribute to speedup without a single approach dominating.

\vspace{-.5ex}
\subsection{Discussion}

This evaluation demonstrates that traditional benchmarks are fundamentally inadequate for assessing workload-driven database optimizations.
Round-robin baselines derived from JOB and CEB produce negligible caching speedups (<1.08$\times$) across all tested systems, suggesting result caching provides minimal value.
However, on the \textit{same systems}, \systemname{} workloads, which closely match production trace characteristics, reveal substantial caching potential: 1.1--2.98$\times$ speedups depending on system architecture and workload variant.

Beyond these aggregate speedups, fine-grained analysis reveals critical architectural trade-offs \rev{visible only} under realistic workload conditions.
Coarse-grained mechanisms such as result caching consistently provide robust benefits with minimal overhead, excelling in read-heavy, high-repetition scenarios.
In contrast, fine-grained mechanisms like predicate, subresult, and scan caching introduce complexity and maintenance costs that can offset their benefits. 
While subresult caching enables partial result reuse in write-interleaved workloads, scan caching often degrades DML performance despite benefits in deterministic baselines.

\noindent\textbf{System-specific differences emerge.} 
For example, System~C benefits primarily from result caching due to workload repetitions, but experiences slowdowns from predicate caching due to invalidation overhead \rev{from} DML operations.
This \rev{shows} System~C excels under high repetition but struggles with write-interleaved patterns.
System~D, in contrast, demonstrates that subresult and scan caching can \rev{yield} additional speedups, with subresult caching outperforming block-level scan caching in mixed workloads, overall \rev{showing} System~D's greater resilience in mixed read-write scenarios.

These findings establish \systemname{} as essential infrastructure for evaluating workload-driven optimizations, exposing performance characteristics and system trade-offs that traditional benchmarks \rev{miss}.
Our evaluation so far has focused on caching and scan-set pruning, as they exemplify a class of optimizations that directly depend on workload structure and temporal locality.

\noindent\textbf{Beyond caching: new evaluation dimensions.}
While our evaluation focused on caching and scan-set pruning, the same methodology can reveal the behavior of other workload-driven mechanisms.
\systemname{} enables controlled and reproducible studies of optimizations that rely on temporal locality or query repetition.
For instance, predicate and partition pruning could be studied to evaluate how well systems avoid unnecessary data access under correlated predicates and data skew or write churn. 
Adaptive clustering and automatic materialization could be examined to measure how often systems reuse internal layouts or partial results given realistic repetition and invalidation rates.
Incremental maintenance and view reuse would reveal whether cached intermediates survive bursty or diurnal write patterns, and plan caching and adaptive re-optimization could show how often plans are reused or recompiled as parameters drift.
Workload-aware cache compaction and eviction policies can be analyzed to understand how systems \rev{decide} which results to retain under memory pressure and repetition bursts.
Finally, learned and adaptive components, such as cost models, caching predictors, or reinforcement-learned strategies, could be tested under temporal drifts instead of synthetic steady-state workloads: questions that traditional benchmarks cannot answer.

\vspace{-3ex}
\section{Conclusion}
This paper introduced \systemname{}, a trace-driven benchmark and workload synthesizer that turns anonymized cloud traces into executable, production-shaped workloads on familiar schemas.
In contrast to static benchmarks, \systemname{} preserves core workload signals including repetitions, temporal patterns with e.g. bursts and idle gaps, and the interleaving of reads (\SELECT) with writes (\INSERT~/~\UPDATE~/~\DELETE).
These signals are crucial for evaluating modern, workload-aware optimizations.

\systemname{} contributes a practical methodology with two complementary synthesis modes. 
The matching-based mode reuses vetted queries from existing benchmarks and orders them to match trace properties, providing a robust and low-risk solution. 
The generation-based mode creates novel SQL guided by trace statistics and schema knowledge to better match structural and runtime proxies (e.g., scansets, join shapes, bytes read) at the cost of higher complexity and potential system errors (e.g., OOMs caused by large intermediate result sizes). 
Together, these modes let practitioners trade off simplicity and fidelity.

Our evaluation on real traces (Redset) and a real-world dataset (IMDb) shows that \systemname{} faithfully reproduces repetition and temporal patterns, including DML-aware effects that bound realistic reuse opportunities.
Across four commercial cloud data warehouses, \systemname{} exposes optimization benefits that traditional benchmarks miss: under cloud-like repetition and read/write interleaving, enabling result/predicate reuse yields substantial speedups (in some cases exceeding 3$\times$), whereas standard benchmarks suggest negligible gains.
A naive baseline based on traditional benchmarks (JOB \& CEB) failed to outline these effects, underscoring the need for trace-shaped workloads.

We believe \systemname{} enables more relevant and reproducible systems research by making it easy to test optimizations under production-like characteristics without access to proprietary workloads or data. 
\systemname{} bridges the gap between synthetic and real workloads by preserving intrinsic workload signals from cloud traces, providing a practical foundation for analyzing and advancing workload-driven optimizations in modern cloud data warehouses.
We release code and artifacts to foster adoption and community contributions.

\begin{acks}
\small
This research was partially funded by the state of Hesse as part of the NHR program (NHR4CES), the LOEWE Spitzenprofessur of the state of Hesse (III 5-519/05.00.003-(0005)), the project "hessian.AI AISC" (Grant No. 01IS22091) by the Federal Ministry of Education and Research (BMBF), as well as by the Deutsche Forschungsgemeinschaft (DFG, German Research Foundation) under Germany's Excellence Strategy (EXC3057/1 “Reasonable Artificial Intelligence”, Project No. 533677015). 
We also thank DFKI Darmstadt and hessian.AI for their support.
\end{acks}

\balance

\bibliographystyle{ACM-Reference-Format}
\bibliography{bib}

\end{document}